\documentclass[aps,floats]{revtex4}
\usepackage{amsmath,amssymb}
\usepackage{graphicx,epsfig}
\usepackage[greek,english]{babel}

\begin{document}
\bibliographystyle {plain}

\def\oppropto{\mathop{\propto}} 
\def\opsimeq{\mathop{\simeq}}
\def\opoverderline{\mathop{\overline}}
\def\operarrow{\mathop{\longrightarrow}}
\def\opsim{\mathop{\sim}}

\def\fig#1#2{\includegraphics[height=#1]{#2}}
\def\figx#1#2{\includegraphics[width=#1]{#2}}


\title{ Many-Body-Localization Transition :  \\
strong multifractality spectrum for matrix elements of local operators } 


\author{ C\'ecile Monthus }
 \affiliation{Institut de Physique Th\'{e}orique, 
Universit\'e Paris Saclay, CNRS, CEA,
91191 Gif-sur-Yvette, France}

\begin{abstract}
For short-ranged disordered quantum models in one dimension, the Many-Body-Localization is analyzed via the adaptation to the Many-Body context [M. Serbyn, Z. Papic and D.A. Abanin, PRX 5, 041047 (2015)] of the Thouless point of view on the Anderson transition : the question is whether a local interaction between two long chains is able to reshuffle completely the eigenstates (Delocalized phase with a volume-law entanglement) or whether the hybridization between tensor states remains limited (Many-Body-Localized Phase with an area-law entanglement). The central object is thus the level of Hybridization induced by the matrix elements of local operators, as compared with the difference of diagonal energies. The multifractal analysis of these matrix elements of local operators is used to analyze the corresponding statistics of resonances. Our main conclusion is that the critical point is characterized by the Strong-Multifractality Spectrum $f(0 \leq \alpha \leq 2)=\frac{\alpha}{2}$, well known in the context of Anderson Localization in spaces of effective infinite dimensionality, where the size of the Hilbert space grows exponentially with the volume. Finally, the possibility of a delocalized non-ergodic phase near criticality is discussed.

\end{abstract}

\maketitle

\section{ Introduction} 

In disordered interacting quantum systems, the Many-Body-Localization Transition (see the recent reviews \cite{revue_huse,revue_altman} and references therein) can be seen as a phase transition for the energy levels and for the excited eigenstates.
In the Ergodic phase where the Eigenstate Thermalization Hypothesis (E.T.H.) \cite{deutsch,srednicki,nature,mite,rigol} holds, the energy levels are governed by the Wigner-Dyson statistics, while the eigenstates display the volume-law entanglement with a prefactor given by the thermal entropy.
In the Many-Body-Localized (MBL) phase, the energy levels are governed by the Poisson statistics,
while the eigenstates display an area-law entanglement \cite{bauer,kjall,alet,alet_dyn,luitz}
allowing in particular efficient tensor networks representations \cite{pekker1,pekker2,friesdorf,sondhi,tensor}. The Strong Disorder Real-Space RG approach  
(see \cite{strong_review,refael_review} for reviews)
developed initially by Ma-Dasgupta-Hu \cite{ma_dasgupta}
 and Daniel Fisher \cite{fisher_AF,fisher,fisherreview}
to construct the ground states of random quantum spin chains,
has been extended for the whole Many-Boy-Localized phase into the Strong Disorder 
RG procedure for the unitary dynamics \cite{vosk_dyn1,vosk_dyn2},
and into the RSRG-X procedure
 to construct the whole set of excited eigenstates 
 \cite{rsrgx,rsrgx_moore,vasseur_rsrgx,yang_rsrgx,rsrgx_bifurcation}.
This construction is possible 
only because the Many-Body-Localized phase is characterized by
an extensive number 
of emergent localized conserved operators \cite{emergent_swingle,emergent_serbyn,emergent_huse,emergent_ent,imbrie,serbyn_quench,emergent_vidal,emergent_ros,emergent_rademaker}. As a consequence, the Strong Disorder RG
 breaks down when the MBL transition towards delocalization is approached, and other types of real-space RG have been proposed for the critical point, based on the notion of insulating or thermalizing blocks of various sizes \cite {vosk_rgentanglement,vasseur_resonant,huse_pasteall} and are in favor of a direct transition towards the ergodic phase. Then, as a consequence of the strong-subadditivity property of the entanglement entropy \cite{grover}, the critical point itself has to satisfy the E.T.H. property, i.e. the entanglement entropy has to coincide with the thermal entropy \cite{grover}. However, another possibility is that the transition is towards a delocalized non-ergodic phase, i.e. the entanglement entropy satisfies the volume law, but not with the thermal coefficient fixed by the E.T.H. \cite{grover,harrisMBL}.
This delocalized non-ergodic scenario is usually explained within
 the point of view that the MBL transition is somewhat similar
to an Anderson
 Localization transition in the Hilbert space of 'infinite dimensionality'
 as a consequence of the exponential growth of the size of the Hilbert space 
with the volume \cite{levitov,gornyi_fock,vadim,us_mblaoki,luca,gornyi,kravtsov_nonergomf}. Note however that the existence of a non-ergodic delocalized phase
remains very controversial even for Random-Regular-Graphs
 that are locally tree-like without
boundaries, as shown by the two very recent studies with opposite conclusions
\cite{mirlin_ergo,altshuler_nonergo}.
In summary, whereas the Many-Body-Localized Phase seems nowadays well understood, there remains open issues concerning both the nature of the critical point and the nature of the delocalized phase (ergodic or non-ergodic) close to the critical point. 

To characterize the MBL transition, Reference \cite{papic} has proposed recently
to adapt to the Many-Body context the Thouless point of view for the Anderson transition :
  the question is whether a local interaction
between two long chains is able to reshuffle completely the eigenstates or not \cite{papic}.
This yields that the statistical properties of the matrix elements of local operators
play an essential role \cite{papic}, as also shown by the adaptation of the Dyson Brownian motion to Many-Body-Localization models \cite{serbyn,c_dysonBM}. These studies have also suggested that multifractality is the good framework to describe these statistical properties \cite{papic,serbyn,c_dysonBM}. The goal of the present work is thus to analyze in details
the problem of the coupling between two longs chains
via the multifractal formalism.

The paper is organized as follows.
In Section \ref{sec_thouless}, the coupling of two long chains via
some local operator is described to introduce the notations.
In Section \ref{sec_matrix}, the statistical properties of the matrix elements of
local operators are analyzed via the multifractal formalism. 
In Section \ref{sec_ratio}, the consequences for the
multifractal statistics of the Hybridization Ratios are derived.
The conclusions are summarized in section \ref{sec_conclusion}.

\section{ Analysis of the local coupling between two large blocks }

\label{sec_thouless}

\subsection{ Effects of the local coupling in the tensor-basis }

Let us consider a disordered quantum spin chain model with nearest-neighbor interactions,
with two independent large blocks of length $N$ : 
the block A contains $N$ quantum spins $i=1,..,N$ and $M=2^{N}$ energy levels $a_i$,
while the block B contains $N$ quantum spins $j=N+1,..,2N$ and $M=2^{N}$ energy levels $b_i$,
with the corresponding spectral decomposition
\begin{eqnarray}
H_A = \sum_{i=1}^{M} a_i \vert  a_i > <  a_i \vert
\nonumber \\
H_B = \sum_{j=1}^{M} b_j \vert  b_j > <  b_j \vert
\label{hA}
\end{eqnarray}
 The level spacing in each block is given at exponential level by the inverse of the size of the Hilbert space
\begin{eqnarray}
\Delta_A = \Delta_B =\frac{1}{M} = 2^{-N}
\label{lsA}
\end{eqnarray}

Let us now introduce the local coupling between the spins $i=N$ and $J=N+1$ 
that are at the boundaries of the two blocks. To be more concrete, let us consider
the explicit case of the coupling via the local magnetization
\begin{eqnarray}
V^{AB} = J_{AB} \sigma_N^x \sigma_{N+1}^x 
\label{vab}
\end{eqnarray}
The full Hamiltonian 
\begin{eqnarray}
H=H_A+H_B+V^{AB}
\label{htot}
\end{eqnarray} 
with an Hilbert space of size ${\cal M}=M^2= 2^{2N}$, i.e. with the level spacing
\begin{eqnarray}
\Delta_{AB} = \frac{1}{\cal M } = \frac{1}{ M^2 } 
\label{levelspacing}
\end{eqnarray}
can be analyzed in the tensor-basis with $i=1,..,M$ and $j=1,..,M$.
\begin{eqnarray}
\vert \phi_{ij} > \equiv \vert  a_i > \otimes \vert  b_j >
\label{tensor}
\end{eqnarray}
via the matrix elements
\begin{eqnarray}
< \phi_{i'j'} \vert H \vert \phi_{ij} > &&   =
(a_i+b_j) \delta_{i,i'} \delta_{j,j'} + V^{AB}_{i'j',ij}
\label{hfullmatrix}
\end{eqnarray}
where the matrix elements of the coupling $V^{AB}$
\begin{eqnarray}
  V^{AB}_{i'j',ij} && \equiv < \phi_{i'j'} \vert V^{AB} \vert \phi_{ij} >
= J_{AB} <  a_{i'} \vert \sigma_N^x \vert a_i >
<  b_{j'} \vert \sigma_{N+1}^x  \vert  b_j > 
\label{hmatrix}
\end{eqnarray}
factorizes into matrix elements of the local operators $\sigma_N^x $ and $\sigma_{N+1}^x $
in each block.

\subsection{ Hybridization ratio between two tensor-states }

The level of hybridization between two tensor states $(ij \ne i'j')$
is determined by the ratio 
\begin{eqnarray}
R_{i'j',ij} && \equiv \left\vert \frac{V^{AB}_{i'j',ij} }{\Delta E_{i'j',ij} } \right\vert
\label{ratioR}
\end{eqnarray}
between the off-diagonal matrix element $V^{AB}_{i'j',ij} $ of Eq. \ref{hmatrix}
and the difference between the diagonal terms $ \Delta E_{i'j',ij}$ of Eq. \ref{hfullmatrix}
\begin{eqnarray}
 \Delta E_{i'j',ij} && \equiv < \phi_{i'j'} \vert H \vert \phi_{i'j'} >
- < \phi_{ij} \vert H \vert \phi_{ij} >    =
(a_{i'}+b_{j'}+ V^{AB}_{i'j',i'j'})- (a_{i}+b_{j}+ V^{AB}_{ij,ij})
\label{deltaE}
\end{eqnarray}
as follows : when the ratio is small $R_{i'j',ij} \ll 1 $, 
the hybridization between the two tensor states $(ij \ne i'j')$ is small,
 when the ratio is large $R_{i'j',ij} \gg 1 $, 
the hybridization between the two tensor states $(ij \ne i'j')$ is large.

\subsection{ Set of hybridization ratios for a given tensor-state  }

For a given tensor-state $
\vert \phi_{i_0j_0} > \equiv \vert  a_{i_0} > \otimes \vert  b_{j_0} >
$
one thus needs to consider the hybridization ratios $R_{i_0j_0,ij} $
with all other tensor states $\vert \phi_{ij} > $
in the Hilbert space of size ${\cal M}=M^2$.
The question of the Many-Body-Localization is
related to the statistics of these ${\cal M}$ 
 hybridization ratios $R_{i_0j_0,ij} $ in the limit of large size 
${\cal M} \to + \infty$.
In this limit, it is convenient to decompose the hybridization ratios into
 two classes
\begin{eqnarray}
{\cal M} = {\cal N}_{mixing}({\cal M})+ {\cal N}_{nonmixing}({\cal M})
\label{2classes}
\end{eqnarray}
according to their behavior in the thermodynamic limit
 ${\cal M} \to + \infty$ : ${\cal N}_{nonmixing} $ represents the number of 
hybridization ratios that converge towards zero $R \to 0$,
while ${\cal N}_{mixing} $ represents the number of 
hybridization ratios that do not converge to zero
(they can diverge $R \to +\infty$ or remain finite $R=O(1)$).

In the Many-Body-Localized phase, where 
the entanglement between the two blocks satisfies the
one-dimensional area-law and thus remains finite,
an eigenstate $\vert \psi> $ of the full Hamiltonian $H=H_A+H_B+V^{AB}$ 
can be well approximated by a Schmidt decomposition involving only
a finite sum of tensor product states.
This means that only a finite number
 of hybridization ratios $R_{i_0j_0,ij} $ do not converge towards zero
\begin{eqnarray}
 {\cal N}^{Loc}_{mixing }({\cal M}\to +\infty )&& \simeq O(1)
\label{mixingloc}
\end{eqnarray}

In the Many-Body-Delocalized phase, where the entanglement follows the volume law, an eigenstate $\vert \psi> $ of $H=H_A+H_B+V^{AB}$ involves a diverging number of nearly equivalent terms
in the Schmidt decomposition, i.e. a diverging number of tensor states 
$\vert \phi_{ij} > $. This requires a diverging number of hybridization ratios
that do not converge towards zero
\begin{eqnarray}
{\cal N}^{Deloc}_{mixing}({\cal M}) && \simeq  {\cal M}^{\rho} 
\label{mixingdeloc}
\end{eqnarray}
with some exponent $\rho>0$.

In summary, the behavior of the entanglement Schmidt
 decomposition of an eigenstate
of $H=H_A+H_B+V^{AB}$ is directly related to the statistics
of the hybridization ratios $R_{i_0j_0,ij} $ of Eq. \ref{ratioR},
 and thus to the statistics
of the matrix elements of local operators 
appearing in the off-diagonal matrix elements
 $V^{AB}_{i_0j_0,ij} $ of Eq. \ref{hmatrix}.
To analyze their statistics in the thermodynamic limit ${\cal M}\to +\infty $,
it is convenient to use the multifractal formalism as described in the next section.

\section{  Multifractal statistics for Matrix elements of local operators  }

\label{sec_matrix}

\subsection{ Local magnetization as an example of local operator }

For the block $A$ containing
$N$ sites with an Hilbert space of size $M=2^N$,
the matrix elements of the magnetization $\sigma_N^z$  
\begin{eqnarray}
m_{nm} &&  \equiv  < a_n\vert \sigma_N^z \vert a_m>
\label{mni}
\end{eqnarray}
are of random sign. To characterize their modulus, 
it is thus convenient to introduce the corresponding Edwards-Anderson matrix 
\begin{eqnarray}
q^{EA}_{nm} && 
 \equiv   \vert < a_n\vert \sigma_N^z \vert a_m>\vert^2= < a_n\vert \sigma_N^z \vert a_m> 
< a_m\vert \sigma_N^z \vert a_n> 
\label{qeanmloci}
\end{eqnarray}
that has the nice property to be doubly stochastic \cite{c_dysonBM},
 i.e. it is a square matrix of size $M \times M$
of non-negative real numbers,
where the sums over any row or any column is unity
\begin{eqnarray}
\sum_{n=1}^M q^{EA}_{nm}  = 1 = \sum_{m=1}^M q^{EA}_{nm} 
\label{bistochloci}
\end{eqnarray}
as a consequence of the completeness identity
\begin{eqnarray}
\sum_{n=1}^M \vert \phi_n> < \phi_n\vert = {\rm Id }
\label{completude}
\end{eqnarray}
and the Pauli matrix identity $(\sigma_i^z)^2={\rm Id } $.
Doubly stochastic matrices appear very often in quantum mechanics
 (see for instance \cite{rigol,louck,luck}). 
The normalization of Eq. \ref{bistochloci} means that the $m=1,2,..,M$ values $q^{EA}_{nm} $
can be interpreted as $M$ random weights normalized to unity,
so that the multifractal formalism is very appropriate to characterize 
their statistics as we now describe.

\subsection{ Multifractal exponents $\tau(q)$ }

After its introduction in fluid dynamics in order to
characterize the statistical properties of turbulence 
(see the book \cite{frisch} and references therein),
the notion of multifractality has turned out to be relevant in many areas of physics
(see for instance \cite{halsey,Pal_Vul,Stan_Mea,Aha,Meakin,harte,duplantier_houches}),
in particular at critical points of disordered models
like Anderson localization transitions \cite{janssenrevue,mirlinrevue}
or in random classical spin models
\cite{Ludwig,Jac_Car,Ols_You,Cha_Ber,Cha_Ber_Sh,PCBI,BCrevue,Thi_Hil,cnancy}.
More recently, the wavefunctions of manybody quantum systems 
have been analyzed via multifractality for ground states in pure quantum spin models
\cite{jms2009,jms2010,jms2011,moore,grassberger,atas_short,atas_long,luitz_short,luitz_o3,luitz_spectro,luitz_qmc,jms2014,alcaraz,c_gsmultif}, as well as for
excited states in MBL models
 \cite{luca_mbl,alet,fradkin,santos}.

Here, we wish to analyze
the statistics of the matrix elements $< \phi_n\vert \sigma_N^z \vert \phi_m> $
where $n$ is fixed and where $m=1,2,..,M$ goes over the whole Hilbert space
of size $M=2^N$ for the block A
via the multifractal formalism as follows. The generalized moments
 $Y_q(M) $ define the multifractal exponents $\tau(q)$
\begin{eqnarray}
Y_q(M) \equiv  \sum_{m=1}^M
 \vert < \phi_n\vert \sigma_N^z \vert \phi_m>\vert^{2q} \oppropto_{M \to +\infty}  M^{- \tau(q)}
\label{IPR}
\end{eqnarray}
The related R\'enyi entropies involve the generalized fractal dimensions $D(q)$
 \begin{eqnarray}
S_q(M) \equiv  \frac{ \ln Y_q(M) }{1-q} \oppropto_{M \to +\infty} D(q) \ln M
\label{renyi}
\end{eqnarray}
with the simple relation
 \begin{eqnarray}
D(q) = \frac{ \tau(q)}{q-1 }
\label{Dqtauq}
\end{eqnarray}
For $q=1$, Eq. \ref{renyi} corresponds to the standard Shannon entropy
 \begin{eqnarray}
S_1(M) \equiv - \sum_{m=1}^M \vert < \phi_n\vert \sigma_N^z \vert \phi_m> \vert^{2}
 \ln \vert < \phi_n\vert \sigma_N^z \vert \phi_m> \vert^{2}
\oppropto_{M \to +\infty} D(1) \ln M
\label{shannon}
\end{eqnarray}

\subsection{ Multifractal spectrum $f(\alpha)$ }

Among the $M$ eigenstates, the number ${\cal N}_M(\alpha)$ of eigenstates $m$
that have with $n$ a matrix element of order
 $\vert < \phi_n\vert \sigma_N^z \vert \phi_m> \vert^2 \propto M^{-\alpha}$
defines the multifractal spectrum $f(\alpha)$
\begin{eqnarray}
{\cal N}_M(\alpha) \oppropto_{M \to \infty} M^{f(\alpha)}
\label{nlalpha}
\end{eqnarray}
The saddle-point calculus in $\alpha$ of the moments of Eq. \ref{IPR}
\begin{equation}
Y_q(M) \simeq \int d\alpha \ M^{f(\alpha)} \ M^{- q \alpha} \oppropto M^{-\tau(q)}
\label{saddle}
\end{equation}
yields 
\begin{eqnarray}
 - \tau(q) = {\rm max}_{\alpha} \left( f(\alpha) - q \alpha \right)
\label{legendrec}
\end{eqnarray}
i.e. $\tau(q)$ and $f(\alpha)$ are related via the Legendre transform
\begin{eqnarray}
   \tau(q)+f(\alpha) && = q \alpha 
\nonumber \\
   \tau'(q) && =  \alpha 
\nonumber \\
f'(\alpha) && = q 
\label{legendre}
\end{eqnarray}

\subsection{ Special values of $q$  }

\subsubsection{ Typical exponent $\alpha_0$ for $q=0$}

For $q=0$, Eq. \ref{IPR} simply measures the size of the Hilbert space $Y_0(M)=M $
corresponding to $\tau(0)=-1$ and $D(0)=1$ and to $f(\alpha_0)=1$ :
 this means that there exists an extensive
number $O(M)$ of states having a matrix element 
$\vert < \phi_n\vert \sigma_N^z \vert \phi_m> \vert^2 \propto M^{-\alpha_0}$,
so that 
\begin{eqnarray}
   \alpha_0 = - \frac{1}{M} \frac{\sum_{m=1}^M  \ln \vert < \phi_n\vert \sigma_N^z \vert \phi_m> \vert^{2}} {  \ln M }  \equiv \alpha^{typ}
\label{alphazero}
\end{eqnarray}
 represents the typical exponent.

\subsubsection{ Shannon exponent $\alpha_1$ for $q=1$}

For the Shannon value $q=1$, 
 the double stochasticity property corresponds to the normalization
\begin{eqnarray}
Y_{q=1}(M) \equiv  \sum_{m=1}^M \vert < \phi_n\vert \sigma_N^z \vert \phi_m>\vert^2 =1
\label{IPRq1}
\end{eqnarray}
so that $\tau(q=1)=0$ and
\begin{eqnarray}
   \alpha_1   = f(\alpha_1) = D(1)
\label{alphaun}
\end{eqnarray}
is called the 'information dimension' that characterizes
 the Shannon entropy of Eq. \ref{shannon}. 

\subsubsection{ Smallest exponent $\alpha_{+\infty}$ for $q \to +\infty$}

In the limit $q \to + \infty$, Eq. \ref{IPR} is dominated by
the configurations having the biggest possible matrix element $ \vert  < \phi_n\vert \sigma_N^z \vert \phi_m> \vert_{max}$ , and thus 
the smallest exponent $\alpha_{+\infty}$ 
\begin{eqnarray}
   \alpha_{+\infty} = - \frac{ \ln \vert  < \phi_n\vert \sigma_N^z \vert \phi_m> \vert_{max}^{2}} { \ln M }  = D(+\infty)
\label{alphaqinfty}
\end{eqnarray}
leading to $Y_{q \to +\infty} \simeq  M^{- q \alpha_{+\infty}} $, since
the corresponding singularity spectrum vanishes $f(\alpha_{+\infty})=0$.
Note that here we consider the typical multifractal spectrum satisfying $f(\alpha) \geq 0$,
that describes the statistics that are found in a typical sample
(and not the average multifractal spectrum $f^{av}(\alpha)$ that keeps some meaning
in the negative region $f^{av}(\alpha) <0 $ in order to describe events that happen in rare samples, see \cite{mirlinrevue} for more details in the context of the Anderson transitions).

\subsection{ Simple examples of multifractal spectra  }

As a comparison, one should keep in mind that 
the monofractal spectrum corresponds to 
\begin{eqnarray}
f^{Mono}(\alpha) = \delta(\alpha-1)
\nonumber \\
\tau^{Mono}(q) = (q-1) 
\label{fmono}
\end{eqnarray}

\subsubsection{ Example of the Weak Multifractality Gaussian spectrum}

The Weak Multifractality spectrum depending
 on the small parameter $\epsilon$ 
is a small deformation of the monofractal limit of Eq. \ref{fmono}
\begin{eqnarray}
f^{Weak}(\alpha) = 1 - \frac{(\alpha-(1+\epsilon) )^2}{4 \epsilon}
\label{fweak}
\end{eqnarray}
So it is Gaussian around the typical value $\alpha_0=1+\epsilon$
and corresponds to the exponents
\begin{eqnarray}
\tau^{Weak}(q) = (q-1) (1- \epsilon q)
\label{tauqweak}
\end{eqnarray}
This Weak Multifractality spectrum appears in particular at Anderson Localization Transition
just above $d=2$ \cite{mirlinrevue}.

\subsubsection{ Example of the Strong Multifractality linear spectrum }

The Strong Multifractality linear spectrum
\begin{eqnarray}
f^{Strong}(\alpha) = \frac{\alpha}{2} \ \ \ {\rm for} \ \  0 \leq \alpha \leq 2
\label{fstrong}
\end{eqnarray}
corresponds to  the exponents
\begin{eqnarray}
\tau^{Strong}(q) && = (2q-1) \ \ \ {\rm for} \ \  q \leq \frac{1}{2} 
\nonumber \\
 && =  0 \ \ \ \ \  \ \  \ \ \  \ \   {\rm for} \ \  q \geq \frac{1}{2} 
\label{tauqstrong}
\end{eqnarray}
This Strong Multifractality spectrum appears in particular at Anderson Localization Transition
in the limit of infinite dimension $d \to +\infty$ \cite{mirlinrevue}
or in related long-ranged power-law hoppings in one-dimension 
 \cite{levitov1,levitov2,levitov3,levitov4,mirlin_evers,fyodorov,fyodorovrigorous,oleg1,oleg2,oleg3,oleg4,olivier_per,olivier_strong,olivier_conjecture,us_strongmultif}.

In the following, we will find that this Strong Multifractality Spectrum 
shows up again. The physical reason is that the MBL transition in one-dimension 
is somewhat similar
to the Anderson Localization transition in the Hilbert space whose geometry is of 'infinite dimensionality' like trees, as a consequence of the exponential growth $M=2^N$ 
of the size of the Hilbert space
with the length $N$, as discussed in \cite{levitov,gornyi_fock,vadim,us_mblaoki,luca,gornyi}.

\subsection{ Matrix elements of local operators in the Delocalized phase }

Let us first consider the case of completely 
random eigenfunctions in the Hilbert space,
as in the Gaussian Random Matrix Ensemble. 
Then the matrix elements 
of the local magnetization display all the same scaling
(only the numerical prefactor changes between diagonal and off-diagonal results
\cite{c_dysonBM})
\begin{eqnarray}
\vert < \phi_n\vert \sigma_N^z \vert \phi_m> \vert^{2} \propto \frac{1}{M}
\label{qdiagdelocm2}
\end{eqnarray}
and correspond to
the simple monofractal spectrum of Eq. \ref{fmono}
\begin{eqnarray}
f^{GOE}(\alpha)= f^{mono}(\alpha) =\delta(\alpha-1)
\label{delocfmultif}
\end{eqnarray}

As recalled in the introduction, another possibility
 is that the delocalized phase is non-ergodic
and described by some spectrum $f^{deloc}(\alpha)$
 different from Eq. \ref{delocfmultif}
(as found for Anderson localization on trees \cite{luca,us_loctreemultif}
or as in some matrix model \cite{kravtsov_nonergomf}).
One general constraint is however that the minimal exponent has to be strictly positive
\begin{eqnarray}
\alpha^{deloc}_{q=+\infty} >0 
\label{delocorigin}
\end{eqnarray}
i.e. the matrix element of a local operator cannot remain finite in the delocalized
phase.

\subsection{Matrix elements of local operators in the Many-Body-Localized phase  }

For Many-Body-Localized eigenfunctions characterized
 by some finite localization length $\xi$,
the matrix elements $\vert < \phi_n\vert \sigma_N^z \vert \phi_m> \vert^2$
with a given state $ \vert \phi_n >$ are expected to be :

(i) finite and of order $2^{-\xi} $ for a finite number $2^{\xi}$ 
of other eigenstates $m$
(i.e. each region of length $\xi$ can be considered as ergodic
within its reduce Hilbert space of size $2^{\xi} $ ). 
In the multifractal language, this means that 
the multifractal spectrum begins at the origin
\begin{eqnarray}
\alpha^{loc}_{q=+\infty} =0 = f(\alpha^{loc}_{q=+\infty}=0)
\label{locorigin}
\end{eqnarray}
in contrast to Eq. \ref{delocorigin} concerning the delocalized phase.

(ii) exponentially small in $N$ for all other $(M-2^{\xi}) $ eigenstates 
\cite{papic}, i.e. they decay as a power-law with respect to the size
$M=2^N$ of the Hilbert space. To obtain a more precise description, 
one needs to know the number of eigenstates $M^{f^{loc}(\alpha)}$
that have a matrix element of order $\vert < \phi_n\vert \sigma_N^z \vert \phi_m> \vert^2 \propto M^{-\alpha}$, i.e. one needs a multifractal description.
Here it is important to stress the difference with the Anderson localization in finite dimension $d$, where the exponential localization decay $e^{-L/\xi}$ is not able to compete with the power-law behavior of the level spacing $\Delta E =1/L^d$. This is why in Anderson localization models, the multifractality appears only at the critical point. 
Here on the contrary,
 the multifractal formalism is needed not only at criticality but also
 in the Many-Body-Localize phase.

\section{  Multifractal analysis of the Hybridization Ratios  }

\label{sec_ratio}

\subsection{ Multifractal statistics of the coupling $ V^{AB}_{i_0j_0,ij} $  }

The coupling of Eq. \ref{hmatrix} directly inherits 
 the multifractal statistics
of the matrix-elements $<  a_{i_0} \vert \sigma_N^x \vert a_i > $ 
and $<  b_{j_0} \vert \sigma_{N+1}^x  \vert  b_j >  $
within each of the two blocks.
Indeed the generalized moments read 
in terms of the size of the global Hilbert space ${\cal M}=M^2$
\begin{eqnarray}
 \sum_{i=1}^{M}\sum_{j=1}^{M} \vert  V^{AB}_{i_0j_0,ij} \vert^{2q} 
&&
=  J_{AB}^{2q} 
\left( \sum_{i=1}^{M}\vert <  a_{i'} \vert \sigma_N^x \vert a_i > \vert^{2q} \right)
\left(\sum_{j=1}^{M} \vert <  b_{j'} \vert \sigma_{N+1}^x  \vert  b_j > \vert^{2q} \right)
\nonumber \\
&& \propto   J_{AB}^{2q}  M^{-\tau(q)}  M^{-\tau(q)} = J_{AB}^{2q}{\cal M}^{-\tau(q)}
\label{Vtauq}
\end{eqnarray}

So the number of tensor states $(i,j)$ that have a 
coupling with the reference state $(i_0,j_0)$ 
scaling as $ \vert  V^{AB}_{i_0j_0,ij} \vert \sim {\cal M}^{-\alpha/2}$
is described by the same multifractal spectrum $f(\alpha)$ of Eq. \ref{nlalpha}
\begin{eqnarray}
{\cal N} ( \vert V^{AB}_{i_0j_0,ij} \vert \sim {\cal M}^{-\alpha/2} ) \propto {\cal M} ^{f(\alpha)} 
\label{falphacoupling}
\end{eqnarray}

\subsection{ Hybridization Ratio between two consecutive energy levels }

Following Ref. \cite{papic}, let us first discuss the case of two
consecutive levels separated by the level spacing 
\begin{eqnarray}
\Delta E_{i_0j_0,next} = \frac{1}{\cal M } 
\label{levelspacingnext}
\end{eqnarray}
The coupling between these two levels is 
governed by the typical exponent $\alpha_{q=0}$ (Eq. \ref{alphazero})
in our present multifractal formalism
\begin{eqnarray}
 \vert V^{AB}_{i_0j_0,next} \vert \sim {\cal M }^{-\alpha_0/2}  
\label{typ}
\end{eqnarray}
So the Hybridization ratio between two consecutive levels scales as
\begin{eqnarray}
\frac{\vert V^{AB}_{i_0j_0,next} \vert}{\Delta E_{i_0j_0,next}} 
= {\cal M }^{1-\alpha_0/2}  
\label{rationext}
\end{eqnarray}

This scaling 
is actually representative of all states
having an energy difference scaling as
$\Delta E_{i_0j_0,i,j} \propto \frac{1}{\cal M }  $.
So the Many-Body-Localized phase can be stable only
if Eq. \ref{rationext} decays with the scale.
This leads to the following bound for the typical
exponent
\begin{eqnarray}
 \alpha_0^{loc} >2
\label{alphatyplocbound}
\end{eqnarray}
Using ${\cal M }=M^2=2^{2N} $, the exponent $\kappa$ characterizing
the Many-Body-Localized phase in Ref. \cite{papic} corresponds to
\begin{eqnarray}
\kappa = (\alpha_0^{loc}-2) \ln 2 
\label{kappa}
\end{eqnarray}
 in our present multifractal description.

In the Delocalized phase, the Hybridization ratio between two consecutive levels
of Eq. \ref{rationext} is expected to diverge, leading to
\begin{eqnarray}
 \alpha_0^{deloc} <2
\label{alphatypdelocbound}
\end{eqnarray}
In particular in the GOE monofractal case of Eq. \ref{delocfmultif} where
\begin{eqnarray}
\alpha_0^{GOE}=1
\label{alphatypgoe}
\end{eqnarray}
the Ratio of Eq. \ref{rationext} diverges as
\begin{eqnarray}
R^{GOE}_{ab,next} =  {\cal M }^{1/2}  
\label{rationextgoe}
\end{eqnarray}

\subsection{ Hybridization Ratio with the state having the biggest off-diagonal coupling }

Another interesting special case is the state $(i,j)$ having
the biggest off-diagonal coupling with the state $(i_0,j_0)$,
which is governed by the exponent $\alpha_{q=\infty}$ of Eq. \ref{alphaqinfty}
in our present multifractal formalism
\begin{eqnarray}
V_{max} = {\rm max}_{(ij)} ( \vert V^{AB}_{i_0j_0,ij}  \vert) \sim  {\cal M }^{- \frac{\alpha_{\infty}}{2}}  
\label{Vmax}
\end{eqnarray}
The energy difference $ \Delta E$ between these two levels 
is finite of order $ {\cal M}^{0}$,
 so that the
Hybridization Ratio keeps the same behavior as Eq. \ref{Vmax}
\begin{eqnarray}
R= \frac{V_{max} }{\Delta E}  \sim {\cal M }^{- \frac{\alpha_{\infty}}{2}}  
\label{max}
\end{eqnarray}

In the Many-Body-Localized phase, this Hybridization Ratio
with the maximally-coupled state is expected to be finite 
in agreement with Eq. \ref{locorigin}
\begin{eqnarray}
\alpha^{loc}_{q=+\infty} =0 = f(\alpha^{loc}_{q=+\infty}=0)
\label{locoriginbis}
\end{eqnarray}

\subsection{ Multifractal statistics of the Hybridization Ratio $R_{i_0j_0,ij}$  }

After the two extreme cases just described, let us now discuss 
the general case where the two levels 
are separated by ${\cal M }^x$ intermediate levels with $0 \leq x \leq 1$.
Taking into account the level spacing $\Delta E=\frac{ 1}{\cal M }  $,
 the energy difference scales as
\begin{eqnarray}
\Delta E_{i_0j_0,ij} = \frac{ {\cal M}^x }{\cal M } = \frac{1 }{ {\cal M}^{1-x} }
\label{levelspacingx}
\end{eqnarray}
So the value $x=0$ corresponds to the level-spacing scaling
(Eq. \ref{levelspacingnext}),
while the value $x=1$ corresponds to finite energy difference.

With a numbering $n=1,..,{\cal M}$ of energy levels,
the change of variables $n={\cal M}^x$ yields $dn = (\ln {\cal M}){\cal M}^x dx$,
i.e. at leading order without the logarithmic factor,
 the number of levels corresponding to a given exponent $x$ is ${\cal M}^x dx $.
From the multifractal statistics of the coupling 
 $V_{i_0j_0,ij}={\cal M }^{-\alpha/2}$, one thus obtains
that the number of Hybridization ratios $R_{i_0j_0,ij}=V_{i_0j_0,ij}/\Delta E_{i_0j_0,ij}$
scaling as $R_{i_0j_0,ij} \sim {\cal M}^{-\frac{\gamma}{2}}$
reads at the levels of exponential factors
\begin{eqnarray}
{\cal N} ( R_{i_0j_0,ij} \sim {\cal M}^{-\frac{\gamma}{2}}  ) 
&& \propto
\int_0^{+\infty} d \alpha  {\cal M}^{f(\alpha)-1}  \int_0^1 dx {\cal M}^x
 \delta ( \gamma- \alpha+2(x- 1) )
\nonumber \\
&&  \propto 
\int_0^{+\infty} d \alpha  {\cal M}^{f(\alpha)-1}  \int_0^1 dx {\cal M}^x \delta ( x- (1-\frac{\alpha-\gamma}{2})  )
\nonumber \\
&&  \propto {\cal M}^{\frac{\gamma}{2}}
\int_0^{+\infty} d \alpha  {\cal M}^{f(\alpha)-\frac{\alpha}{2}} \theta(\gamma \leq \alpha \leq \gamma+2) \equiv {\cal M}^{F(\gamma)}
\label{NR}
\end{eqnarray}
that defines the multifractal spectrum $F(\gamma) $ 
associated to the Hybridization Ratio for $-2 \leq \gamma \leq +\infty$.

\subsection{ Number ${\cal N}_{mixing} $ of levels with a non-vanishing Hybridization Ratio} 

Using the notation of Eq. \ref{2classes},
one obtains that the number of 
hybridization ratios that do not converge towards zero
corresponds to the region $-2 \leq \gamma  \leq 0$
\begin{eqnarray}
{\cal N}_{mixing}({\cal M}) && =
 \int_{-2}^0 d \gamma {\cal N} ( R \sim {\cal M}^{-\frac{\gamma}{2}}  ) 
\nonumber \\
&& \propto \int_{-2}^0 d \gamma  {\cal M}^{\frac{\gamma}{2}}
\int_0^{+\infty} d \alpha  {\cal M}^{f(\alpha)-\frac{\alpha}{2}} \theta( \gamma \leq \alpha \leq \gamma+2)
\nonumber \\
&& \propto 
\int_0^{2} d \alpha  {\cal M}^{f(\alpha)-\frac{\alpha}{2}} 
\int_{-2+\alpha}^0 d \gamma  {\cal M}^{\frac{\gamma}{2}}
\label{classRinfty}
\end{eqnarray}

Since the last integral is dominated by the boundary $\gamma=0$, one obtains
\begin{eqnarray}
{\cal N}^{mixing} && \simeq \int_0^{2} d \alpha  {\cal M}^{f(\alpha)-\frac{\alpha}{2}}
 = {\cal M}^{F(\gamma=0)}
\label{nres}
\end{eqnarray}
in terms of the value $F(\gamma=0)$ at $\gamma=0$
of the multifractal spectrum of Eq. \ref{NR}.

In Eq. \ref{nres},
 one has to do the same saddle-point calculation as in Eq. \ref{saddle}
for $q=1/2$ except for the domain of integration,
so it is important to discuss where the saddle point $\alpha_{q=1/2}$ is
within the region $[0,2]$.

\subsection{ Delocalized phase  }

In the Delocalized phase, using the bounds of Eqs \ref{delocorigin}
and \ref{alphatypdelocbound},
one obtains that the saddle point
$\alpha_{q=1/2}$ lies in the integration interval of Eq. \ref{nres}
\begin{eqnarray}
0< \alpha^{deloc}_{q=+\infty} \leq \alpha^{deloc}_{q=1/2} \leq \alpha^{deloc}_{q=0} <2
\label{interdeloc}
\end{eqnarray}
 Then Eq. \ref{saddle}
yields using Eq. \ref{Dqtauq}
\begin{eqnarray}
{\cal N}_{mixing}^{deloc} && \simeq  {\cal M}^{f(\alpha_{q=1/2})-\frac{\alpha_{q=1/2}}{2}}
=   {\cal M}^{-\tau(q=1/2)} ={\cal M}^{\frac{D(q=1/2)}{2}} 
\sim {\cal M}^{F^{deloc}(\gamma=0)}
\label{nmixingdeloc}
\end{eqnarray}
i.e. the growth of the number of mixing levels
is governed by the dimension $D^{deloc}(q=1/2)$ corresponding to $q=1/2$,
and this dimension has thus to be positive in the delocalized phase
(Eq \ref{mixingdeloc})
\begin{eqnarray}
F^{deloc}(\gamma=0) = \frac{D^{deloc}(q=1/2)}{2} =f^{deloc}(\alpha_{q=1/2})-\frac{\alpha_{q=1/2}}{2}  >0
\label{Fdeloc}
\end{eqnarray}
The point $\alpha_{q=1/2} \in ]0,2[ $ maximizing $(f(\alpha)-\frac{\alpha}{2} ) $
on the interval $0<\alpha<2$
 satisfies the inequality
\begin{eqnarray}
f^{deloc}(\alpha_{q=1/2}) > \frac{\alpha_{q=1/2}}{2}
\label{Fdelocineq}
\end{eqnarray}

For instance for the GOE case where $D^{GOE}(q=1/2)=1 $
 (Eq. \ref{delocfmultif}), one obtains the scaling
\begin{eqnarray}
{\cal N}_{mixing}^{GOE} && \simeq {\cal M}^{\frac{1}{2}} = M
\label{nmixingGOE}
\end{eqnarray}
which coincides with the maximal number of terms in the Schmidt decomposition
of an eigenstate. In addition, plugging the monofractal spectrum 
$f^{GOE}(\alpha)=\delta(\alpha-1)$ into Eq. \ref{NR}
yields 
\begin{eqnarray}
{\cal N}^{GOE} ( R \sim {\cal M}^{-\frac{\gamma}{2}}  ) \equiv {\cal M}^{F^{GOE} (\gamma)}
&&  \propto {\cal M}^{\frac{\gamma+1}{2}}
 \theta(-1 \leq \gamma \leq 1 ) 
\label{NRgoe}
\end{eqnarray}
so that the spectrum of hybridization ratios reads
\begin{eqnarray}
F^{GOE} (\gamma) = \frac{\gamma+1}{2} \theta(-1 \leq \gamma \leq 1 ) 
\label{NRgoespec}
\end{eqnarray}

\subsection{ Many-Body-Localized phase  }

In the Many-Body-Localized phase, the 
number of mixing ${\cal N}^{mixing} $ of Eq. \ref{nres} has to remain finite 
(Eq \ref{mixingloc})
\begin{eqnarray}
F^{loc}(\gamma=0) = 0
\label{Floc}
\end{eqnarray}
In addition, the saddle-point evaluation of Eq. \ref{nres}
should be entirely due to the origin $\alpha=0=f^{loc}(0)$ (Eq. \ref{locorigin})
as a consequence of the effective Hilbert space of size $2^{\xi}$.
This requirement that the saddle-point of Eq. \ref{nres} is at the origin
means that
\begin{eqnarray}
\alpha^{loc}_{q=1/2} = 0
\label{alloc}
\end{eqnarray}
and that the multifractal spectrum for non-vanishing $\alpha$
satisfies the strict inequality
\begin{eqnarray}
f^{loc}(\alpha) < \frac{\alpha}{2} \ \ \ \ \ {\rm for } \ \ \  0 < \alpha \leq 2
\label{falphacritiloc}
\end{eqnarray}
In particular this yields that on the interval $0 < \alpha \leq 2 $,
the multifrctal spectrum remain strictly below the value unity $1=f(\alpha_0)$
corresponding to the typical exponent $\alpha_0$
\begin{eqnarray}
f^{loc}( 0 < \alpha \leq 2) <  1 = f^{loc}(\alpha_0)
\label{falphacritilocbound}
\end{eqnarray}
i.e. one recovers that the typical exponent $\alpha_0$ should 
satisfy the bound $\alpha_0^{loc}>2$ of Eq. \ref{alphatyplocbound}.

Taking into account
 the concavity property of the multifractal spectrum $f(\alpha)$
between the origin $\alpha=0=f(\alpha=0)$ and the typical value $f(\alpha_0)=1$
for $\alpha= \left(1-\frac{\alpha}{\alpha_0} \right) 0+ \frac{\alpha}{\alpha_0} \alpha_0$
\begin{eqnarray}
f^{loc}(  \alpha ) \geq \left(1-\frac{\alpha}{\alpha_0} \right) f(0)+ \frac{\alpha}{\alpha_0} f(\alpha_0) = \frac{\alpha}{\alpha_0} \ \ \ \ \ {\rm for } \ \ \  0 \leq \alpha \leq \alpha_0
\label{concavity}
\end{eqnarray}
and the bound of Eq. \ref{falphacritiloc}, one obtains that near the transition
when the typical exponent is close to $\alpha_0 = 2+\epsilon$,
there is no room left for curvature, and the multifractal spectrum
is constrained to be linear of the form
\begin{eqnarray}
f^{loc}( \alpha ) \simeq \frac{\alpha}{\alpha_0} \ \ \ \ \ {\rm for } \ \ \  0 \leq \alpha \leq \alpha_0=2+\epsilon
\label{locnearcriti}
\end{eqnarray}
 Note that this linear spectrum for the Localized phase
has already been found for the MBL case in \cite{c_mblperturb}
 and for an Anderson Localization matrix model in \cite{kravtsov_nonergomf}.

Plugging the linear spectrum of Eq. \ref{locnearcriti} into Eq. \ref{NR}
yields
\begin{eqnarray}
{\cal N}^{loc} ( R \sim {\cal M}^{-\frac{\gamma}{2}}  ) \equiv {\cal M}^{F^{loc}(\gamma)}
&&  \propto {\cal M}^{\frac{\gamma}{2}}
\int_0^{+\infty} d \alpha  {\cal M}^{ \left(\frac{1}{\alpha_0}-\frac{1}{2}\right) \alpha} \theta(\gamma \leq \alpha \leq \gamma+2) 
 \theta ( 0 \leq \alpha \leq \alpha_0=2+\epsilon )
\label{NRloclin}
\end{eqnarray}
Since $\alpha_0>2$, the integral in $\alpha$ is dominated by 
the smallest value of the integration interval, which is $\alpha_{min}=\gamma$ 
for $0<\gamma<\alpha_0$, and which is $\alpha=0$ for $-2<\gamma<0$, so that one obtains
the following spectrum for the hybridization ratio $R$
\begin{eqnarray}
F^{loc}(\gamma) && \simeq \frac{\gamma}{\alpha_0} \ \ \ \ \ {\rm for } \ \ \  0 \leq \gamma \leq \alpha_0=2+\epsilon
\nonumber \\
F^{loc}(\gamma) && \simeq \frac{\gamma}{2} \ \ \ \ \ {\rm for } \ \ \  -2 \leq \gamma \leq 0
\label{NRloclingamma}
\end{eqnarray}
Since the second lign corresponds
to the negative region $F^{loc} (-2 \leq \gamma <0 ) <0 $, 
it only represents rare events that do not occur in a typical sample,
the typical multifractal spectrum is given by the first line alone,
which remarkably coincides with the spectrum of matrix elements of Eq. \ref{locnearcriti}.

\subsection{ Critical point  }

As the critical point is approached from the Many-Body-Localized phase 
$\epsilon= \alpha_0-2 \to 0^+$, the multifractal spectrum of Eq. \ref{locnearcriti}
becomes
\begin{eqnarray}
f^{criti}(\alpha) = \frac{\alpha}{2} \theta(0 \leq \alpha \leq 2)
\label{falphacriticriti}
\end{eqnarray}
that coincides with the known Strong Multifractality spectrum of Eq. \ref{fstrong}
with the exponents
\begin{eqnarray}
D^{criti}(q) && =  \frac{ 1-2q }{1-q} \theta( q \leq \frac{1}{2} )
\label{tauqcriti}
\end{eqnarray}
The vanishing dimensions for $q \geq 1/2$ and in particular for $q=1/2$
\begin{eqnarray}
D^{criti}(q=\frac{1}{2}) && =  0
\label{tauqcritidemi}
\end{eqnarray}
matches the delocalized result of Eq. \ref{nmixingdeloc}.

The difference with the Many-Body-Localized phase
is that the integral of Eq. \ref{nres} for the number of mixing levels
has now contributions from the whole interval $0 \leq \alpha \leq 2$
(instead of the concentration at the origin $\alpha=0$), i.e. the resonances
do not correspond only to finite matrix elements described by the value $\alpha=0$,
but involve all possible decays of matrix elements  $\vert V^{AB} \vert \sim {\cal M }^{-\alpha/2}  $ with $0 \leq \alpha \leq 2$.

For the hybridization Ratio, the multifractal spectrum of Eq. \ref{NR}
 reads at criticality
\begin{eqnarray}
F^{criti} ( \gamma ) && = \frac{\gamma}{2} \theta( -2 \leq \gamma \leq 2) 
\label{Fgammacriti}
\end{eqnarray}
where the negative region $F^{criti} ( \gamma ) <0 $ for $\gamma<0$
can only describe rare events that do not occur in a typical sample,
while the positive region $F^{criti} ( \gamma ) \geq 0 $ that describes
what happens in a typical sample
\begin{eqnarray}
F^{criti}_{typ} ( \gamma ) && = \frac{\gamma}{2} \theta( 0 \leq \gamma \leq 2) 
\label{Fgammacritipos}
\end{eqnarray}
 coincides with the Strong Multifractal spectrum of Eq. \ref{falphacriticriti}.

\subsection{ Is a delocalized non-ergodic phase possible ? }

As recalled in the Introduction, the existence of a delocalized non-ergodic phase
remains very controversial. If it exists, it should be characterized by
a multifractal spectrum intermediate between the critical spectrum
 of Eq. \ref{falphacriticriti} and the monofractal GOE spectrum $f^{GOE}(\alpha)=\delta(\alpha-1)$.
Above, we have obtained that in the Many-Body-Localized phase and at criticality,
there are enough mathematical and physical 
requirements to determine the multifractal spectrum.
In the Delocalized phase however, it is not clear to us presently
if it is also possible to constraint the possible spectra.

However it is interesting to analyze the following 
family of linear spectra 
depending on the continuous parameter $\alpha_{\infty} \in [0,1]$
obtained in an Anderson localization matrix model \cite{kravtsov_nonergomf}
\begin{eqnarray}
f^{deloc}_{\alpha_{\infty}}(\alpha) = \left( \frac{\alpha}{2}+\frac{\alpha_{\infty}}{2} \right) 
\theta(\alpha_{\infty} \leq \alpha \leq \alpha_0=2-\alpha_{\infty})
\label{falphadelocnonergo}
\end{eqnarray}
It can be considered as the simplest possible interpolation
 between the critical spectrum of Eq \ref{falphacriticriti}
for $\alpha_{\infty}=0$ and the monofractal ergodic spectrum for $\alpha_{\infty}=1$.
The corresponding generalized dimensions read
\begin{eqnarray}
D^{deloc}_{\alpha_{\infty}}(q) && = \frac{1-q(2-\alpha_{\infty})}{1-q} \ \ \ \ \ {\rm for } \ \ \  q \leq \frac{1}{2}
\nonumber \\
D^{deloc}_{\alpha_{\infty}}(q) && =  \alpha_{\infty} \ \ \ \ \ {\rm for } \ \ \  q \geq \frac{1}{2}
\label{dqdelocnonergo}
\end{eqnarray}
Note that the minimal exponent $\alpha_{\infty}$ describes
all generalized dimensions $D^{deloc}_{\alpha_{\infty}}(q) $ for $q \geq 1/2$
and determines in particular the scaling of the number of mixing levels
of Eq. \ref{nmixingdeloc}
\begin{eqnarray}
{\cal N}_{mixing}^{deloc} && \propto {\cal M}^{\frac{D(q=1/2)}{2}} 
\sim {\cal M}^{\frac{\alpha_{\infty}}{2} }
\label{nmixingdelocnonergo}
\end{eqnarray}

Plugging Eq. \ref{falphadelocnonergo} into Eq \ref{NR}
yields 
\begin{eqnarray}
{\cal N}^{deloc} ( R \sim {\cal M}^{-\frac{\gamma}{2}}  )  \equiv {\cal M}^{F^{deloc} (\gamma)}
&&  \propto {\cal M}^{\frac{\gamma}{2}+\frac{\alpha_{\infty}}{2}}
\int_0^{+\infty} d \alpha  \theta(\gamma \leq \alpha \leq \gamma+2)
\theta(\alpha_{\infty} \leq \alpha \leq \alpha_0=2-\alpha_{\infty})
\nonumber \\
&&  \propto {\cal M}^{\frac{\gamma}{2}+\frac{\alpha_{\infty}}{2}}
  \theta(\alpha_{\infty}-2 \leq \gamma \leq 2-\alpha_{\infty})
\label{NRdelocnonergo}
\end{eqnarray}
so that the corresponding spectrum for the hybridization ratio 
\begin{eqnarray}
F^{deloc} (\gamma) = \left( \frac{\gamma}{2}+\frac{\alpha_{\infty}}{2} \right)
  \theta(\alpha_{\infty}-2 \leq \gamma \leq 2-\alpha_{\infty})
\label{Fdelocnonergo}
\end{eqnarray}
interpolates between the critical result of Eq. \ref{Fgammacriti}
 for $\alpha_{\infty}=0 $ and the GOE result of Eq. \ref{NRgoespec} for $\alpha_{\infty}=1 $. This spectrum is negative for $\gamma<-\alpha_{\infty} $,
so the typical spectrum that describes what happens in a given sample
corresponds to the region where the spectrum is positive
\begin{eqnarray}
F^{deloc}_{typ} (\gamma) = \left( \frac{\gamma}{2}+\frac{\alpha_{\infty}}{2} \right)
  \theta(- \alpha_{\infty} \leq \gamma \leq 2-\alpha_{\infty})
\label{Fdelocnonergotyp}
\end{eqnarray}

In summary, if a delocalized non-ergodic phase exists, we feel that Eq. \ref{falphadelocnonergo} is probably the best candidate. Further work is needed to determine whether it is indeed realized or whether it can be ruled out.
From a numerical point of view, the simplest and clearest criterion should be
the typical exponent $\alpha_0$ satisfying $\alpha_0^{loc}>2= \alpha_0^{criti} > \alpha_0^{deloc}$ : the ergodic phase corresponds to $\alpha_0^{ergo}=1$, while 
a delocalized non-ergodic phase would correspond to 
\begin{eqnarray}
1< \alpha_0^{nonergo}<2
\label{alphatypnonergotyp}
\end{eqnarray}
So the question is whether the typical exponent displays a discontinuous jump
from $\alpha_0^{criti}=2$ to $\alpha_0^{ergo}=1$, or whether it remains continuous
with values in the whole interval of Eq. \ref{alphatypnonergotyp}.

\section{ Conclusion }

\label{sec_conclusion}

In this article, the local coupling between two long disordered quantum spin chains has been analyzed via the multifractal statistics of matrix elements of local operators in order to determine whether the interaction is able to reshuffle completely the eigenstates, as expected in the Delocalized phase with a volume-law entanglement, or whether the hybridization between tensor states remains limited, as expected in the Many-Body-Localized Phase with an area-law entanglement. 

Our main conclusion is that for Many-Body-Localization models in one dimension
with nearest-neighbors couplings, there are actually 
enough mathematical and physical requirements to determine completely the 
multifractal spectrum at criticality : it is found to be the Strong-Multifractality Spectrum $f(0 \leq \alpha \leq 2)=\frac{\alpha}{2}$. This result can be seen as 
the generalization to the whole spectrum of the criterion concerning only the typical exponent $\alpha_0^{typ}=2$ proposed previously
 and tested numerically in Ref. \cite{papic}
(see Eq \ref{kappa} and the corresponding discussion).
At the other end of the spectrum,
the criterion concerning the minimal exponent $\alpha_{q=+\infty}^{loc}=0$
(Eq. \ref{locorigin} and the corresponding discussion) can be related to the
existence of an extensive number of
emergent localized conserved operators in the Many-Body-Localized phase \cite{emergent_swingle,emergent_serbyn,emergent_huse,emergent_ent,imbrie,serbyn_quench,emergent_vidal,emergent_ros,emergent_rademaker}.

This Strong-Multifractality Spectrum is well-known in the context of Anderson Localization in spaces of effective infinite dimensionality, where the size of the Hilbert space grows exponentially with the volume. It should be stressed however that the observable displaying this multifractality is different in the two cases :
in Anderson Localization models, the multifractality describes the inhomogeneities of the wave-function in real space and is measured by the Inverse-Participation-Ratios; in Many-Body-Localization models, the multifractality concerns the matrix elements between Hilbert space eigenstates of a local operator in real space. This shows one again that all the subtleties of the Many-Body-Localization problem are related to the understanding of what happens in the Hilbert space, what happens in real space and the interplay between the two.
The fact that matrix elements of local operators are essential observables for the Many-Body-Localization transition is confirmed by their role in the Dyson Brownian approach
 \cite{serbyn,c_dysonBM} where they govern the repulsion between energy levels.
Physically, this Strong-Multifractality Spectrum represents the most inhomogeneous
possibility within the multifractal point of view, and the corresponding critical statistics of energy levels is the closest possible to the Poisson statistics of the Localized phase, a possibility that has been studied via singular perturbation theory in \cite{c_mblperturb}, where the strong multifractality spectrum was also found to describe the entanglement spectrum at criticality.
Note that this multifractal scenario 
for the Many-Body-Localization transition
is very different from the current Renormalization Group proposals
for the MBL critical point,
which are based on the decomposition of the chain into insulating
 and thermalizing blocks of various sizes
 \cite {vosk_rgentanglement,vasseur_resonant,huse_pasteall}.
The multifractal picture is more similar to the standard description of the
Anderson localization transitions, where the inhomogeneities at criticality are
described by a multifractal spectrum and not by a black-and-white 
decomposition into localized and delocalized blocks.

Finally, we have discussed whether a delocalized non-ergodic phase could exist
and proposed the numerical criterion of Eq. \ref{alphatypnonergotyp}
 to try to clarify this very controversial issue.

\end{document}